%% file: main.tex
\documentclass[sigconf]{acmart}

\usepackage{graphicx}
\usepackage{textcomp}
\usepackage{tcolorbox}
\usepackage{pifont}
\usepackage{xcolor}
\usepackage{soul}
\usepackage{booktabs}
\usepackage{multirow}
\usepackage{float}
\usepackage{url}
\usepackage{balance}
\usepackage{microtype}
\usepackage[numbered]{bookmark}
\usepackage{tcolorbox}
\usepackage{titlesec}
\usepackage{enumitem}
\usepackage{seqsplit}
\usepackage{newunicodechar,graphicx}
\usepackage[utf8]{inputenc}
\usepackage{pgf-pie}  
\usepackage{draftwatermark}
\setlist[itemize,enumerate]{noitemsep, topsep=0pt, leftmargin=1.5em}
\setlist{nolistsep}

\AtBeginDocument{%
  \providecommand\BibTeX{{%
    \normalfont B\kern-0.5em{\scshape i\kern-0.25em b}\kern-0.8em\TeX}}}

\renewcommand\footnotetextcopyrightpermission[1]{}
\hypersetup{
	pdfborder          = {0 0 0},
	breaklinks         = true,
	bookmarksnumbered  = true,
	pdfkeywords        = {software engineering, code quality, test smells, unit testing, education, open-source tool, tsdetect},
	pdftitle           = {Insights from the Field: Exploring Students' Perspectives on Bad Unit Testing Practices},
	pdfauthor          = {Anthony Peruma,Eman Abdullah AlOmar,Wajdi Aljedaani,Christian D. Newman,Mohamed Wiem Mkaouer}
}
\DeclareRobustCommand{\okina}{%
  \raisebox{\dimexpr\fontcharht\font`A-\height}{%
    \scalebox{0.8}{`}%
  }%
}
\newunicodechar{ʻ}{\okina}
\usepackage{listings}

\definecolor{javared}{rgb}{0.6,0,0} 
\definecolor{javagreen}{rgb}{0.25,0.5,0.35} 
\definecolor{javapurple}{rgb}{0.5,0,0.35} 
\definecolor{javadocblue}{rgb}{0.25,0.35,0.75} 
 
\newcommand{\SchoolR}{Rochester Institute of Technology}
\newcommand{\SchoolS}{Stevens Institute of Technology}
\newcommand{\SchoolT}{University of North Texas}

\newcommand{\RQA}{\textbf{RQ1:} How do students perceive the harmful nature of test smells?}

\newcommand{\RQB}{\textbf{RQ2:} To what extent does an IDE plugin assist students in enhancing their understanding of test smells?}

\copyrightyear{2024}
\acmYear{2024}
\setcopyright{acmlicensed}\acmConference[ITiCSE 2024]{Proceedings of the 2024 Innovation and Technology in Computer Science Education V. 1}{July 8--10, 2024}{Milan, Italy}
\acmBooktitle{Proceedings of the 2024 Innovation and Technology in Computer Science Education V. 1 (ITiCSE 2024), July 8--10, 2024, Milan, Italy}
\acmDOI{10.1145/3649217.3653643}
\acmISBN{979-8-4007-0600-4/24/07}

\begin{document}

\title{Insights from the Field: Exploring Students' Perspectives on Bad Unit Testing Practices}

\author{Anthony Peruma}
\email{peruma@hawaii.edu}
\orcid{0000-0003-2585-657X}
\affiliation{%
  \institution{University of Hawaiʻi at Mānoa}
  \state{Hawaiʻi}
  \country{USA}
}

\author{Eman Abdullah AlOmar}
\email{ealomar@stevens.edu}
\orcid{0000-0003-1800-9268}
\affiliation{%
  \institution{Stevens Institute of Technology}
  \state{New Jersey}
  \country{USA}
}

\author{Wajdi Aljedaani}
\email{wajdi.aljedaani@unt.edu}
\orcid{0000-0002-6700-719X}
\affiliation{%
  \institution{University of North Texas}
  \state{Texas}
  \country{USA}
}

\author{Christian D. Newman}
\email{cnewman@se.rit.edu}
\orcid{0000-0002-8838-4074}
\affiliation{%
  \institution{Rochester Institute of Technology}
  \state{New York}
  \country{USA}
}

\author{Mohamed Wiem Mkaouer}
\email{mmkaouer@umich.edu}
\orcid{0000-0001-6010-7561}
\affiliation{%
  \institution{University of Michigan-Flint}
  \state{Michigan}
  \country{USA}
}

\begin{abstract}
Educating students about software testing practices is integral to the curricula of many computer science-related courses and typically involves students writing unit tests. Similar to production/source code, students might inadvertently deviate from established unit testing best practices, and introduce problematic code, referred to as test smells, into their test suites. Given the extensive catalog of test smells, it becomes challenging for students to identify test smells in their code, especially for those who lack experience with testing practices. In this experience report, we aim to increase students' awareness of bad unit testing practices, and detail the outcomes of having 184 students from three higher educational institutes utilize an IDE plugin to automatically detect test smells in their code. Our findings show that while students report on the plugin's usefulness in learning about and detecting test smells, they also identify specific test smells that they consider harmless. We anticipate that our findings will support academia in refining course curricula on unit testing and enabling educators to support students with code review strategies of test code. 
\end{abstract}

%
%



\maketitle

\SetWatermarkText{Preprint} 
\SetWatermarkScale{0.5} 

\section{Introduction}
\label{Section:Introduction}

Unit testing is an industry-standard software engineering technique that involves writing small, self-contained tests to verify the correctness of individual code units. Unit testing is the basic building block of software testing, and therefore the Association for Computing Machinery (ACM) recommends the integration of unit testing in Computer Science and Software Engineering curricula \cite{ACM-Curricula}. Consequently, seeking educational practices for teaching software testing is becoming the focus of various researchers, resulting in a Software Testing Education Workshop \cite{TestEd2023}. 

These education-related studies focus on designing appropriate pedagogical methods of teaching unit test programming and identifying appropriate testing frameworks that can be adopted for educational purposes. Yet, little is known about how to assess the non-functional quality of students' tests, in order to avoid bad programming practices in their test code. Bad programming practices in the test code, also known as \textit{test smells}, are indicators of potential design problems in the test suite. Test smells are similar to code smells (i.e., bad programming practices in production code). The existence of test smells negatively impacts the efficiency and effectiveness system's testing by increasing the flakiness of test cases \cite{Camara2021}, and hindering test code readability and understandability \cite{van2001refactoring}.

Prior research has correlated the prevalence of test smells with developers' deviation from adopting best practices of writing test code, such as X-Unit \cite{meszaros2007xunit}. Therefore, it is critical to raise awareness of test smells among students (i.e., early career developers) and to provide them with the tools and training needed to write high-quality test code. In this study, we aim to determine \textit{what type of test smells are perceived by students to be non-harmful to their test code}. To do so, we designed a two-step experiment in which students initially write unit test cases for an input project. Then, we provide students with \texttt{tsDetect}, a test smell detection tool, to identify smelly units in their test suites \cite{tsdetect}. Originally, \texttt{tsDetect} is a command line tool. However, for this study, we obtained the source code for the tool from its authors and converted the command line tool to an IntelliJ IDE plugin \cite{Peruma_TSDetect_-_IntelliJ}.

Students are instructed to address the smelly test only when they think it is \textit{worth it}, i.e., it is harmful to their test code. Finally, we follow up with a survey to identify the main reasons behind their decisions, particularly when it contradicts the state-of-the-art research findings in this domain.

The results of our study show the existence of a variation between researchers and students on what is considered to be harmful smell types. In particular, the Lazy Test smell was on the top of the list of least harmful smells, followed by Magic Number Test, Eager Test, Empty Test, and Duplicate Assert.

To avoid the propagation of these misperceptions, the findings of our study motivate educators to illustrate the negative aspects of all types of test smells, with a particular focus on how their existence is harmful to test functionality and code comprehension. Also, we show how \texttt{tsDetect}, as an IDE plugin, was found useful by students to improve the overall quality of their test code. 

\section{Related Work}
\label{Section:Related}
The research literature presents a rich body of work on automatic approaches and tools for identifying test smells in software testing \cite{aljedaani2021test}. Investigators have invested substantial effort in identifying and classifying a wide range of test smells \cite{garousi2018smells}. In contrast, some researchers have directed their efforts toward understanding the implications of test smells and formulating effective strategies for their elimination \cite{kim2021secret}. For instance, Van Bladel and Demeyer \cite{van2017test} introduced a technique to eliminate test smells within the domain of refactoring test code. Similarly, Van Deursen et al. have extensively addressed harmful test smells, providing well-founded techniques for their mitigation \cite{van2001refactoring}. Furthermore, they have introduced innovative conceptual and technical frameworks to evaluate students' coding activities by detecting test smells in their codebases, leading to valuable observations and insights related to test smells. These research pursuits have made notable contributions to the comprehension and management of test smells, significantly impacting software testing practices and overall code quality assessment.

Numerous methods \cite{aaltonen2010mutation,edwards2004using} have been devised and explored in evaluating test suites generated by students. In a recent study, Bai et al. \cite{bai2022check} examined the student's effectiveness in utilizing a testing checklist on writing the test case, focusing on aspects related to completeness, effectiveness, and maintainability. Buffardi and Aguirre-Ayal \cite{buffardi2021unit} conducted an investigation into the origins of unit test smells and their relationship with inaccurate test results, employing a thorough examination of students' testing assignments. The study explored the correlation between the test accuracy of students' assignments and the specific types of test smells present in their codebases. In a complementary study, Bai et al. \cite{bai2021students} conducted an experimental investigation to understand students' comprehension of unit testing and the obstacles they encountered. Furthermore, an empirical investigation was conducted by Bavota et al. \cite{bavota2015test}, involving both students and industrial developers, to assess the effect of test smells on software comprehension tasks.

In academic research and educational contexts, scholars and instructors frequently use test suites as an evaluation rate to assess the quality of source code created by the students \cite{bai2019exploring,kazerouni2017quantifying,williams2001support}. However, students' productivity is typically determined by metrics such as lines of code every working hour \cite{baheti2002exploring} or any designated coding session \cite{kazerouni2017quantifying}. Regrettably, the importance of test smells in the educational setting has been overlooked or underemphasized in the existing literature. Bai et al. \cite{bai2022check} conducted a study focusing on the influence of the Assertion Roulette smell on students' productivity and coding behavior, utilizing the Bowling Score Keeper project as an illustrative context. In a similar study, Aljedaani et al. \cite{aljedaani2023test} conducted a controlled experiment carried out involving 96 undergraduate computer science students to examine the influence of two prevalent test smells, specifically "Assertion Roulette" and "Eager Test," on the student's proficiency in debugging and troubleshooting test case failures.

\section{Method}
\label{Section:Method}
\begin{figure*}[t]
 	\centering
 	\includegraphics[trim=0cm 0cm 0cm 0cm,scale=0.6]{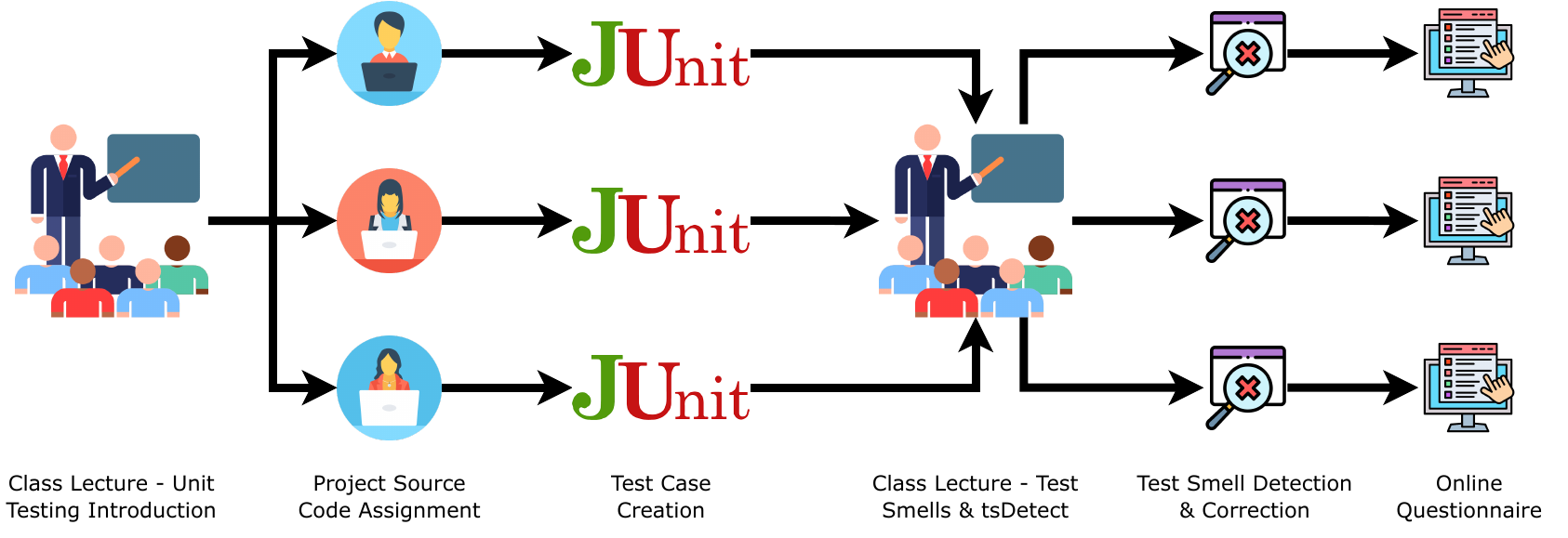}
 	\caption{Overview of the setup of our study.}\vspace{-2mm}
 	\label{Figure:method}
\end{figure*}

In this section, we provide a detailed explanation of our study's research environment, including information on participants, activities, and data collection. Figure \ref{Figure:method} presents a high-level overview of the process, where participants were given a project's source code and tasked with writing test cases. After completing the task, participants were introduced to the concept of test smells and guided on using an IntelliJ IDEA plugin, \texttt{tsDetect}, to detect these smells in their code. Subsequently, they were optionally asked to complete an online questionnaire.

It is important to mention that prior to conducting the study with our student participants, we conducted a pilot run involving a group of three teaching assistants. This pilot study aimed to identify areas for improvement. Valuable feedback from the pilot study allowed us to identify flaws in the assignment source code, and technical challenges related to the plugin,\texttt{tsDetect} setup and usage, along with some ambiguities in the lecture notes and questionnaires. Consequently, we fixed source code errors, created video tutorials for the plugin, and addressed ambiguity and duration issues based on the feedback received. The responses to the questionnaire from the pilot run have been excluded from our analysis.

Furthermore, since this study is conducted at multiple locations, lecture content and activity instructions were shared among institutes to minimize the risk of inconsistencies. 
Finally, to enable replication and extension, lecture artifacts, coding assignments, \texttt{tsDetect}, and survey questions are available at \cite{Artifacts}.  

\subsection{Participants}
To ensure a representative sample, our study recruited participants from three higher education institutions located in the United States: \SchoolR, \SchoolS, and \SchoolT. These participants consisted of both graduate and undergraduate students, the majority of whom were enrolled in a Computing-related major at their respective institute, and were concurrently taking a software engineering-related course taught by one of the authors.  Participants were not provided monetary compensation for their involvement in the study. A total of 190 participants were enrolled in our study. However, six participants did not complete the assignment and the questionnaire and were excluded from our analysis. Therefore, in this study, we report on the experience and perception of 184 participants. Table \ref{Table:participants} provides a breakdown of participants by institute, degree, and primary major type.

\subsection{Coding Activity}
We constructed three Java programs, and each participant was randomly assigned one program. The programs were intentionally designed to strike a balance between simplicity and functionality. They were made simple enough to ensure participants could comprehend their behavior, yet complex enough to allow for the creation of multiple test cases. A summary description of each program is provided below:
\begin{itemize}
    \item Automated Teller Machine - Includes functionality to verify a customer, retrieve a customer's balance, deposit money, withdraw money, change a customer's PIN, and transfer money from one account to another.
    \item Vending Machine - Involves adding and vending items, with necessary checks for item availability and price. 
    \item Calculator - This program contains methods to perform common arithmetic operations on a single or collection of digits  
\end{itemize}
\begin{table}
\centering
\caption{A breakdown of the participants in our study by institute and degree type.}
\label{Table:participants}
\begin{tabular}{@{}lllr@{}}
\toprule
\multicolumn{1}{c}{\textbf{\begin{tabular}[c]{@{}c@{}}Institute\\ Name\end{tabular}}} &
  \multicolumn{1}{c}{\textbf{\begin{tabular}[c]{@{}c@{}}Degree\\ Type\end{tabular}}} &
  \multicolumn{1}{c}{\textbf{\begin{tabular}[c]{@{}c@{}}Primary\\ Major\end{tabular}}} &
  \multicolumn{1}{c}{\textbf{\begin{tabular}[c]{@{}c@{}}Recruited\\ Participants\end{tabular}}} \\ \midrule
\multirow{2}{*}{RIT\footnotemark} & \multirow{2}{*}{Graduate} & Data Science         & 70           \\
                     &                           & Applied Statistics   & 1            \\ \midrule
\multirow{3}{*}{SIT\footnotemark} & Undergraduate             & Software Engineering & 3            \\
                     & \multirow{2}{*}{Graduate} & Software Engineering & 34           \\
                     &                           & Cybersecurity        & 1            \\ \midrule
\multirow{2}{*}{UNT\footnotemark} & Undergraduate             & Computer Science     & 70           \\
                     & Graduate                  & Computer Science     & 5            \\ \midrule
\multicolumn{3}{r}{\textit{\textbf{Total Recruited
Participants}}}                             & \textbf{184} \\ \bottomrule
\end{tabular}
\end{table}
\footnotetext[1]{\SchoolR}
\footnotetext[2]{\SchoolS}
\footnotetext[3]{\SchoolT}

The participants were instructed not to alter the source code. Their task was writing test cases that would achieve a minimum of 85\% code coverage. Furthermore, participants were instructed to use IntelliJ IDEA as their IDE and the JUnit 4 testing framework to construct their test suite. Importantly, the concept of test smells was not introduced or discussed with participants during the initial class lecture on unit testing and was not included in the instructions for the coding activity. This ensured that participants approached the task without prior knowledge or bias related to test smells. The participants had to complete the coding activity within five days and submit their work to the instructor.  

\subsection{Test Smell Detection \& Correction}
After completing the coding activity, the participants had to attend a lecture on test smells. This lecture discussed the different types of test smells and their negative impact on the overall quality of the software system and maintenance activities. Additionally, participants received instructions on how to utilize the \texttt{tsDetect} plugin to identify test smells in their test code.

In the subsequent activity, participants were tasked with running the \texttt{tsDetect} plugin on their test code, written in the previous activity. The participants were instructed to correct the identified test smell instances, as suggested by the plugin, but only if they agreed with its assessment that the reported smell instance needs to be fixed. It was perfectly acceptable if the participants disagreed with specific instances flagged by \texttt{tsDetect}. As before, participants were prohibited from modifying the source code; only modifications to the test code were permitted. This activity had to be completed within five days and submitted to the instructor.  

\subsection{Online Questionnaire}
The final activity in the study involved participants completing an online questionnaire after the test smell detection and correction activity. The questionnaire contained 25 questions, incorporating single-choice, multiple-choice, and open-ended questions. Furthermore, for some single- and multi-choice questions, we asked the participants to explain their choice selection using free text. As part of the analysis, the authors performed a thematic analysis \cite{cruzes2011recommended} of the free text responses to identify recurring themes or patterns within the responses, thereby providing context to better understand their responses. These questions aimed to gather insights on various aspects, including the participants' experience and skill level, perceptions of test smells, and feedback regarding \texttt{tsDetect}.  

\section{Results}
\label{Section:Results}
In this section, we present the findings of the questionnaire responses. We first provide an overview of the participants' general experience, followed by a detailed analysis of their feedback and experiences with the coding activity in our survey. When reporting our findings for RQ1 and RQ2, we incorporate qualitative data from participants' free-text responses to specific questions. This feedback text is presented as written by the participants.

\subsection{General Experience}
While our survey did not collect personally identifiable information, we did gather information on the participants' programming and industry experience. Concerning software engineering industry experience, 89 participants (48.37\%) reported having no industry experience. Moving forward, approximately 92.39\% of the 184 participants had over a year of programming experience, with 50\% having between 3 to 5 years of programming background. However, 48.91\% of all participants had less than one year of Java experience.

As for testing, only 19 participants (10.33\%) had no prior unit testing experience. Additionally, we found that only 25 participants (13.59\%) were unfamiliar with JUnit, and 14 participants (7.61\%) were not acquainted with the IntelliJ IDEA IDE. 

In summary, the above demographic data demonstrates the diversity and representativeness of our sample population; our study's findings are not biased toward a specific group of students, as we have students with varying degrees of experience.

\subsection{\RQA}

\input{assets/pie-chart}

Research on test smells shows that they negatively impact the internal quality of the test suite and hinder the maintainability of the test suite \cite{bavota2015test}. Nevertheless, it is important to note that its negative impact can vary subjectively, depending on the developer and the coding context \cite{Peruma2019}. To this extent, this RQ aims at understanding, from a student's point of view, the types of test smells they consider not harmful. This RQ helps us to better comprehend the subjective nature of test smells and their evaluation, and it helps educators refine curriculum and train future developers effectively.

As part of the questionnaire, we asked participants to indicate and justify what, if any, test smells they consider to be not harmful. Subsequently, the authors examined the participants' responses and identified the prevalent reasons for participants classifying a specific smell type as non-harmful by adopting a thematic analysis approach \cite{cruzes2011recommended}. We performed this analysis to determine the main reasons students thought these smells were not harmful.

As shown in Figure \ref{Figure:Top5HarmfulSmels}, Lazy Test smell is the one most participants (around 19.81\%) regard as not harmful, followed by Magic Number Test, Eager Test, Empty Test, and Duplicate Assert. Conversely, only 2.15\% of participants consider Sensitive Equality, Sleepy Test, Exception Handling, Conditional Test Logic, and Resource Optimism smells as not harmful. Further, approximately 11.69\% of participants considered all the smell types they encountered as harmful.

Next, we analyze participants' reasons for deeming certain smells harmless. Through thematic analysis for each smell type, we learn the rationale behind their views. Below, for each of the top five smells deemed harmless, we elaborate on the common rationales.

\noindent \textbf{Lazy Test.} This smell occurs when multiple test methods test the same source method. The problem with this practice is that it can cause difficulty maintaining consistency and the possibility of duplicating test cases \cite{Bavota2012}. However, in our study, participants believe that such smells do not impact maintenance activities as long as test cases pass and they achieve the desired code coverage. Participants created multiple test methods to test a single source method, where each test method verifies a different part/behavior of the source method, leading to the emergence of this smell. Such practices lead to more focused test cases. As one participant commented, ``The reason for using multiple test cases is to produce high line coverage. The detection of lazy test makes no sense in that case.''

\noindent \textbf{Magic Number Test.} This smell occurs when numeric literals are utilized within assert methods, negatively impacting the test's understandability \cite{Peruma2019}. Participants consider this smell to be unharmful since it does not cause test case failures. Participants feel that this smell is not practical as it requires the addition of more lines of code (i.e., constant declarations), which they consider redundant. As stated by a participant, ``The magic number smell does not necessarily cause any errors in a program. It is just that we need an extra step to elaborate it better.'' Listing \ref{Listing:example1} shows an example.

\begin{lstlisting}[caption={An example of a test case with and without a Magic Number Test smell. Participants regarded this as a non-harmful smell and did not find adding more lines of code to support documented numeric literals practicable.}, label=Listing:example1, firstnumber = last, escapeinside={(*@}{@*)}]
@Test // Magic Number Test smell is present
public void NegSqrt() {
    double result = calculator.squareRoot(-4);
    assertEquals(Double.MIN_VALUE, result, 0);
}

@Test // Magic Number Test smell is not present
public void NegSqrt() {
    double i = -4;
    double result = calculator.squareRoot(i);
    double delta = 0;
    assertEquals(Double.MIN_VALUE, result, delta);
} 
\end{lstlisting}

\noindent \textbf{Eager Test.} This smell occurs when a single test method checks multiple source methods, negatively impacting comprehension and maintenance activities. Similar to other non-harmful smells, participants' feedback shows that this smell does not cause failures. The participants did not see the advantage of having separate test methods for the different source method calls. As one participant stated, ``I don't think the Eager Test is harmful because having multiple test cases within a single test method would be the same as having the test cases in separate test methods.'' However, while this might help cut down the lines of code, it makes maintenance of the test method more challenging, which is a fact participants did not realize due to their lack of experience.

\noindent \textbf{Empty Test.} Having a test method without an assertion method gives rise to this smell, as such test methods will always be reported as passing \cite{Peruma2019}. As any project test codebase grows, empty tests become less visible, misleading testers that a given method in the code is always correct. In this small-scale class assignment, empty tests do not present a major issue resulting in participants not understanding the severity of this smell, as stated by one participant, ``I feel like since it is empty, its like having a variable you dont use.''

\noindent \textbf{Duplicate Assert.} This smell occurs when the same condition is tested multiple times within the same test method, leading to an increase in the length of the test method \cite{Peruma2019}. Participants did not feel that having duplicates of an assertion method was problematic, as it does not affect the test outcomes. One participant stated, ``Duplicate assert is not harmful because it just detects the same assert that we use multiple times.''

A common belief shared by participants when marking a smell type as non-harmful is that since they do not experience test case failures or runtime errors, they consider the smells as not harmful. Participants are primarily concerned about passing test cases and code coverage. As far as participants are concerned, the increase in lines of code due to duplication of testing conditions and source method calls does not impact the maintainability or understandability of the test suite. Furthermore, as test cases pass, some participants regard the detected instances of test smells as false positives (e.g., ``I feel it is a false positive, we call the same method for the test code coverage which should not be an issue/ error'').   
There are a few potential reasons for these outcomes. One reason is that participants may not fully understand the potential negative impacts of these smells, possibly due to a lack of experience (for example, a magic number appearing in a small, easy-to-read program might seem relatively benign). Another reason is that these assignments are designed to introduce concepts and not real-world systems. Consequently, participants might find it challenging to grasp the effort required for comprehending and maintaining the code. Finally, there is also the possibility that some instances detected by \texttt{tsDetect} are, in fact, false positives.

Our findings represent education opportunities. Future research, based on these participants' feedback, can help us understand where education on code quality might be lacking. We can use this to determine what materials are essential for us to improve upon/add to curriculums. We can also understand more about how students' awareness of code quality evolves and how we can support this evolution through tools and education.

\subsection{\RQB}

\input{assets/table-answers}

As described in Section \ref{Section:Method}, this study involves participants using the IDE plugin \texttt{tsDetect} to detect test smells. Hence, this RQ examines participants' feedback about the plugin by examining the plugin's ease of use, including the ease of interpreting the output, the detection accuracy of the plugin, and the overall learning experience. Through this analysis, we gain an understanding of the plugin's performance and its impact on users. These insights serve as input to guide improvements and enhancements to ensure that the plugin evolves to meet the user's needs and expectations better. 

\subsubsection{\textbf{Usability:}}
As shown in Table \ref{Table:anwers01}, the majority of the participants found the plugin easy to use, with 18.48\% providing a neutral response, while only 10.33\% rating the ease of use as difficult. Further, participants had to give a free text explanation for their choice selection. An analysis of the responses shows participants highlighting areas such as installing the plugin, the stability and performance of the plugin, and the plugin's user interface. 

Notably, some participants encountered challenges during the installation process or faced environmental issues. However, once successfully installed, participants found the plugin's user interface intuitive and straightforward. For instance, one participant remarked, ``Setting up the plugin took time and was a little confusing, but once it was set up the rest was easy.'' However, at the same time, we did see a few participants having difficulty performing the activity due to trouble understanding the instructions or their lack of experience in unit testing, and test smells (e.g., ``Hard to understand the purpose of the tool, and what the goal was. ''). Encouragingly, most participants reported no performance degradation while running the plugin, with one participant commenting, "Getting results very faster and easy to analyze."

Moving on, approximately 66.3\% of participants found it easy to interpret the output, while 14.13\% faced difficulty. Once more, analyzing the free text responses, we observe feedback about visualization, documentation, and prior knowledge. 

One common area of improvement is the incorporation of comprehensive documentation into the tool to provide details about the detected smells, including recommendations for fixing them. Currently, the plugin provides only the name of the detected smell, which is insufficient, especially for less experienced users, like most participants in this study. For example, one participant remarked, ``If I had not used the testsmells.org website, I wouldn't have known what to fix just by looking at the code.'' On a positive note, participants appreciated the visualizations provided by the plugin, such as the pie chart and tables. For instance, a participant commented,  `` The pie chart was easy to understand. Additionally, the option of viewing the specific infected class and methods was very helpful.'' 

\subsubsection{\textbf{Detection Accuracy:}}
From Table \ref{Table:anwers01}, we observe that a majority of participants, approximately 80.43\%, expressed satisfaction with the plugin's detection accuracy, while 3.26\% were dissatisfied.

Reviewing the textual answers associated with their choices, we notice feedback associated with false positives and inexperience. Interestingly, the feedback associated with the ``Unsure'' option shares similarities with those who selected ``Dissatisfied.''  Common feedback includes participants indicating the occurrence of false positive smells, as in this example: ``It seemed like a lot of false positives, so I feel mixed.'' However, at the same time, we did encounter instances of participants acknowledging their lack of experience, which hindered their ability to confidently assess the accuracy of the plugin's detections. For instance, one participant mentioned, ``i am unsure how accurate it is especially i need to be super familiar.'' 

\subsubsection{\textbf{Learning Experience:}}
While the purpose of \texttt{tsDetect} is to help developers improve the internal quality and maintainability of their test cases, it also helps educate developers about the concept and types of test smells. To this extent, we asked participants about the learning effectiveness of the plugin. As shown in Table \ref{Table:anwers01}, an overwhelming majority of participants, approximately 94.58\%, stated that the plugin contributed to their improved understanding of test smells to varying degrees. Only a small minority of 10 participants felt that the plugin did not significantly aid them in better grasping the concept of test smells. Finally, when asked if the plugin's suggestions helped improve their code, 91.3\% of participants responded ``Yes'', while the remaining 8.7\% responded ``No''.

\section{Reflections}
\label{Section:Reflections}
In this section, we reflect on our experience of utilizing an IDE plugin to enhance our teaching of test smells for students. While the students' overall feedback was positive, we identified areas that could benefit from improvement. Our reflections are not limited to educators planning on using this or similar tools, but also to the research community and tool vendors/builders.

\subsection{Education}
Engaging students in hands-on activities enhances their overall learning experience by reinforcing the concepts covered in lectures. Participants' feedback shows that \texttt{tsDetect} is a valuable tool for educating students about test smells. As an IDE plugin, \texttt{tsDetect} enables students to seamlessly view both the code and the test smell detection results without switching between multiple applications. For instance, one participant remarked, ``The tool was very easy to use and was helpful in understanding test smells within the test cases I had written,'' while another stated, ``Overall, I just wanted to say that it was really great to learn such a different topics related to testing and it would be really useful in the future as well.''

Educators should emphasize to students that while writing tests to achieve high code coverage is essential, this high coverage should not come at the expense of test code quality. For example, one participant remarked, ``The plugin shows lazy test error even though the code line coverage is 100\%.''
Furthermore, educators should encourage the use of test smell detection tools, such as \texttt{tsDetect}, in addition to source/production code quality tools, such as PMD and linters, to improve the maintainability of test suites. 

Due to the wide array of laptop/desktop configurations and environments, it's likely that some students may face difficulties during the installation and setup of these tools. The time and effort invested by students, educators, and teaching assistants to address these challenges can negatively impact the quality of the students' learning experience. One solution to mitigate such risks is to provide pre-built virtual machines with running instances of these tools.

\subsection{Academic Research}
When compared to prior work that involved student feedback on test smells, our findings show both similar and contrasting findings. More specifically, Bai et al. \cite{Bai2022} reports that the smell Assertion Roulette is not considered harmful by students. However, in our case, approximately 2.86\% of participants consider this smell type as not harmful. In contrast, our work shows similarities with Aljedaani et al. \cite{aljedaani2023test}, where the authors report that students encounter more challenges working with code exhibiting Assertion Roulette than code having the Eager Test smell. These comparisons highlight the varying student views on test smells' harmfulness and contribute to broadening our understanding of software testing education. 

\subsection{Tooling}
Tool vendors/developers should construct tools to accommodate users with diverse levels of experience. For instance, aside from creating tools that efficiently perform their desired functionality, tool developers should also include help guides/documentation to assist users in using the tool. This is particularly crucial in the context of smells (both code and test smells), owing to the multitude of distinct smell types, and challenges that novice users encounter in comprehending the adverse effects of these specific smells and the techniques for correcting them. For instance, as one participant commented, ``It was easy to read the output but if one is not familiar with the type of test smells it was not very useful.''

\section{Conclusion}
\label{Section:Conclusion}
Unit testing is an essential practice in software development that involves writing code to test individual components or units of source/production code to verify their correctness and functionality. As such, writing high-quality and maintenance-friendly test code is essential and should be emphasized when teaching students about software testing. In this paper, we report on our experience of utilizing a test smell detection IDE plugin, \texttt{tsDetect}, to complement our teaching of test smells to undergraduate and graduate students in three higher education institutes. Our findings show that, while using \texttt{tsDetect} helps students understand the concept of test smells through hands-on activities, there are also challenges with using this tool. Additionally, we report on specific test smells that students report as not harmful, such as the Eager Test smell.

\bibliographystyle{ACM-Reference-Format}
\bibliography{main}

\end{document}

%% file: assets/pie-chart.tex
\begin{figure}
\centering
\begin{tikzpicture}
\pie{19.81/Lazy Test,
    12.17/Magic Number Test,
    7.40/Eager Test,
    4.53/Empty Test,
    4.06/Duplicate Assert,
    11.69/None,
    40.33/Other 15 Smells}  
\end{tikzpicture}\vspace{-3.5mm}
\caption{Distribution of (top 5) test smells considered \textbf{\textit{not harmful}} by student participants.}
\vspace{-3mm}
\label{Figure:Top5HarmfulSmels}
\vspace{-2mm}
\end{figure}
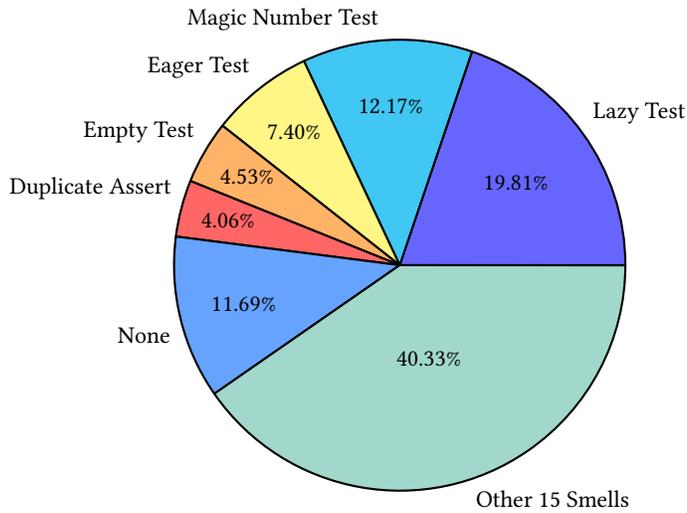

%% file: assets/table-answers.tex
\begin{table*}[t]
\centering
\caption{Answers to questions examining participants' feedback about the test smell detection IDE plugin, \texttt{tsDetect}.}
\vspace{-3mm}
\label{Table:anwers01}
\resizebox{\textwidth}{!}{%
\begin{tabular}{@{}lrr|lrr|lrr|lrr@{}}
\toprule
\multicolumn{3}{c|}{\textit{\textbf{\begin{tabular}[c]{@{}c@{}}How would you rate \\ the plugin's ease of use?\end{tabular}}}} &
  \multicolumn{3}{c|}{\textit{\textbf{\begin{tabular}[c]{@{}c@{}}Interpreting the output generated \\ by the plugin was:\end{tabular}}}} &
  \multicolumn{3}{c|}{\textit{\textbf{\begin{tabular}[c]{@{}c@{}}Rate your level of satisfaction with the \\ detection accuracy of the plugin:\end{tabular}}}} &
  \multicolumn{3}{c}{\textit{\textbf{\begin{tabular}[c]{@{}c@{}}To what extent did the plugin help \\ you better understand test smells?\end{tabular}}}} \\ \midrule
\multicolumn{1}{c}{\textbf{Answer Options}} &
  \multicolumn{1}{c}{\textbf{Count}} &
  \multicolumn{1}{c|}{\textbf{Percentage}} &
  \multicolumn{1}{c}{\textbf{Answer Options}} &
  \multicolumn{1}{c}{\textbf{Count}} &
  \multicolumn{1}{c|}{\textbf{Percentage}} &
  \multicolumn{1}{c}{\textbf{Answer Options}} &
  \multicolumn{1}{c}{\textbf{Count}} &
  \multicolumn{1}{c|}{\textbf{Percentage}} &
  \multicolumn{1}{c}{\textbf{Answer Options}} &
  \multicolumn{1}{c}{\textbf{Count}} &
  \multicolumn{1}{c}{\textbf{Percentage}} \\ \midrule
Very easy to use &
  70 &
  38.04\% &
  Easy &
  89 &
  48.37\% &
  Satisfied &
  119 &
  64.67\% &
  The plugin helped me understand test smells to a moderate extent &
  59 &
  32.07\% \\
Somewhat easy to use &
  61 &
  33.15\% &
  Neutral &
  36 &
  19.57\% &
  Unsure &
  30 &
  16.3\% &
  The plugin helped me understand test smells to a great extent &
  59 &
  32.07\% \\
Neutral &
  34 &
  18.48\% &
  Very easy &
  33 &
  17.93\% &
  Very satisfied &
  29 &
  15.76\% &
  The plugin somewhat helped me better understand test smells &
  34 &
  18.48\% \\
Somewhat difficult to use &
  17 &
  9.24\% &
  Difficult &
  22 &
  11.96\% &
  Dissatisfied &
  6 &
  3.26\% &
  The plugin greatly helped me understand test smells &
  22 &
  11.96\% \\
Very difficult to use &
  2 &
  1.09\% &
  Verry difficult &
  4 &
  2.17\% &
  Very dissatisfied &
  0 &
  0\% &
  The plugin did not help me better understand test smells &
  10 &
  5.43\% \\ \bottomrule
\end{tabular}%
}
\end{table*}